\PassOptionsToPackage{colorlinks,citecolor=blue,urlcolor=black,linkcolor=blue}{hyperref}
\documentclass[10pt]{iopart}

\expandafter\let\csname equation*\endcsname\relax
\expandafter\let\csname endequation*\endcsname\relax

\usepackage{makecell}
\usepackage{soul}
\usepackage{nicematrix}
\usepackage{graphicx}
\usepackage{amsmath}
\usepackage{mathrsfs}
\usepackage{tabularx}
\usepackage{amsmath}
\usepackage{orcidlink}
\usepackage{hyperref}
\usepackage{subcaption}
\usepackage{multirow}
\usepackage{longtable}
\usepackage{booktabs}
\usepackage{siunitx}

\usepackage[backend=biber, style=ieee, eprint=false]{biblatex}
\AtEveryBibitem{%
  \ifentrytype{online}{%
  }{%
    \clearfield{howpublished}%
    \clearfield{url}%
    \clearfield{urldate}%
    \clearfield{issn}%
    \clearfield{related}%
  }%
}

\usepackage{math-functions}

\AddToHookNext{shipout/foreground}{%
  \begin{tikzpicture}[remember picture,overlay]
    \node[anchor=south west, draw=black, line width=0.4pt,
      inner sep=6pt, outer sep=0pt,
      text width=\dimexpr\textwidth-12pt\relax,
      align=left, font=\normalfont\footnotesize]
      at ([xshift=\dimexpr1in+\hoffset+\oddsidemargin\relax,
           yshift=47mm]current page.south west)
      {This work has been submitted to a journal for possible publication.
       Copyright may be transferred without notice, after which this version
       may no longer be accessible.};
  \end{tikzpicture}%
}

\renewcommand{\arraystretch}{1.25}

\begin{document}

\title[]{Fixed-mesh based approach for modeling of superconducting magnetic bearings}

\author{Elias Paakkunainen\textsuperscript{1,2,*}, Bárbara Maria Oliveira Santos\textsuperscript{3}, Gabriel dos Santos\textsuperscript{4}, Timo Tarhasaari\textsuperscript{2}, Paavo Rasilo\textsuperscript{2}, Sebastian Schöps\textsuperscript{1}}
\address{$^{1}$Computational Electromagnetics Group, TU Darmstadt, Germany \\ $^{2}$Electrical Engineering Unit, Tampere University, Finland \\ $^{3}$Rio de Janeiro State University, Brazil \\ $^{4}$Universidade Federal do Rio de Janeiro, Brazil}
\ead{elias.paakkunainen@tuni.fi }
\hspace{73pt}{\footnotesize \textsuperscript{*}Corresponding author.}
\begin{abstract}
Numerical simulations of superconducting magnetic bearings (SMBs) are complicated by the relative motion between high temperature superconductors and the magnetic guideway, which commonly requires repeated updates of the geometry and mesh. We present a fixed-mesh based approach for modeling the motion induced effects in which a coordinate transformation is applied only to the air region between the moving regions. The transformation can be included in a time-dependent anisotropic material tensor, leaving the mesh during the simulation and the rest of the formulation unchanged. To provide numerical examples, the approach is implemented in 2D using the finite element method with the $j$-$a$ formulation and is combined with homogenization and circuit coupling for double-crossed loops of coated conductors. The open-source model implementations are made available. The proposed approach is verified and validated against existing modeling approaches and experimental results from the literature for SMBs. Two simulation examples are presented, and in both of them the levitation force predicted by the fixed-mesh model agrees with the experimental results, yielding $R^2$ values above 0.99. For a verification problem, it is shown that the fixed-mesh model converges towards the remeshed reference solution under mesh refinement as expected. For an SMB undergoing vertical and horizontal translations, the homogenized fixed-mesh model reproduces the lateral-force response of a fully resolved moving-geometry reference model. Additionally, the homogenized fixed-mesh model is computationally more efficient than the reference model, indicating a significant reduction in the number of degrees of freedom and the computation time. The results demonstrate accurate and computationally efficient simulation of translational motion with the fixed-mesh based approach. Because the motion is modeled through a modified material tensor, the proposed approach can be easily applied to existing simulation workflows and to other finite element formulations.

\end{abstract}

\noindent{\it Keywords}: Finite element method, fixed-mesh, high temperature superconductors, homogenization, magnetic levitation, motion, superconducting magnetic bearings.

\maketitle

\section{Introduction}
\label{sec:introduction}
The use of magnetic levitation (MagLev) systems has been proposed for public transportation~\cite{bernstein_superconducting_2020}. There are three main technologies for MagLev systems: electrodynamic suspension, electromagnetic levitation, and superconducting levitation. The first two technologies are highly dependent on control systems to guarantee levitation, whereas superconducting levitation offers naturally stable levitation~\cite{bernstein_superconducting_2020}. Several large-scale prototypes have been developed, including the first man-loaded superconducting MagLev, in China~\cite{wang_first_2002}, the SupraTrans project in Germany~\cite{Schultz_2005} and the MagLev Cobra project in Brazil~\cite{sotelo_maglev_2014}. The levitating element in each superconducting magnetic levitation train is a cryostat where the superconducting material is placed, be it high temperature superconducting (HTS) bulks~\cite{wang_first_2002,Schultz_2005,sotelo_maglev_2014} or stacks of HTS coated conductors (CCs)~\cite{Martins_2019aa,Santos_2023aa}.

The development of this technology not only includes novel experimental techniques, but also novel simulation techniques, capable of modeling the superconducting magnetic bearings (SMBs) in full detail. Often, the finite element method (FEM) is applied, with different electromagnetic formulations, such as the $h$~\cite{Sass_2015aa,queval_superconducting_2018,Grilli_2018aa}, $t$-$a$~\cite{Santos_2023aa} and $h$-$\phi$~\cite{Saske_2025aa} formulations, to model the electromechanical behavior of the superconductors as they interact with the magnetic guideway and levitate. This approach allows the full assessment of the behavior of the SMB operation. 

In FEM, spatial discretization of the examined geometry (e.g. full 3D system or a cross-section) yields a mesh which is used for the computations. Modeling relative motion between superconductors and the magnetic guideway can be challenging, especially due to the coupling of the meshes in the moving regions (which contain the moving bodies). A commonly used approach is to remesh the finite element~(FE) model at each time step, or when the movement exceeds a certain threshold. This approach leads to the additional computational cost of remeshing, and to potential numerical problems as the solution needs to be interpolated between meshes~\cite{Yamazaki_1997aa}. Alternatively, the field generated by the magnetic guideway can be imposed on the HTS as a boundary condition which changes over time to recreate the effect of motion~\cite{queval_superconducting_2018}. This approach does not require remeshing, but the coupling between the HTS and the magnetic guideway is unidirectional, i.e., the currents induced in the HTS do not affect the magnetic field generated by the guideway. Another approach from the literature suggests a coupled boundary element method and FEM model which eliminates to need to mesh the air region~\cite{Kurz_1998aa}. This method is applied to a levitator system with normal conducting materials in~\cite{Kurz_1998aa}.

\begin{figure}
    \centering
    \includegraphics[width=0.5\linewidth, trim={43.5bp 50bp 42bp 57.75bp}, clip]{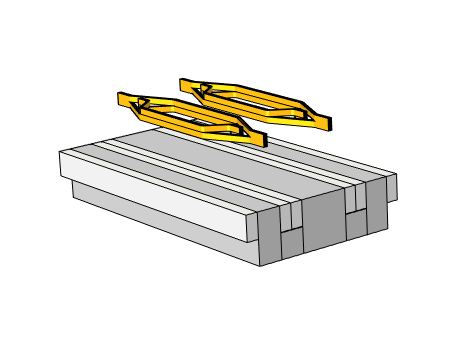}
    \caption{3D representation of the copper-colored double-crossed loops of coated conductors over a magnetic guideway, in gray.}
    \label{fig:dcl_geometry}
\end{figure}

In this work, the motion of the model regions is represented by modifying the material properties in the space between the moving parts. In the literature, similar approach has been proposed in modeling of rotating electrical machines with FEM~\cite{Kettunen_2014aa}. In our modeling setting, the quantity which is eventually capturing the effects of the motion is the reluctivity $\nu$ of the air region between the moving parts. The mathematical background of this approach is in the invariance of the electromagnetic fields on the chosen system of coordinates. This technique is applied to a superconducting magnetic levitator composed of two sets of double-crossed loops (DCLs) of CCs~\cite{Martins_2019aa,Santos_2023aa} levitating over a magnetic guideway in the same format as the one used in MagLev Cobra, see figure~\ref{fig:dcl_geometry}. The goal is to reduce the complexity, particularly in terms of setting up the model, of this type of simulation, providing simpler and potentially faster FE simulations of superconducting moving bodies.

The rest of the paper is structured as follows. In section~\ref{sec:model}, we present the numerical methods used in this work, including the $j$-$a$ FE formulation. In section~\ref{sec:results}, numerical examples to verify the proposed approach and to demonstrate its capabilities are shown. Finally, section~\ref{sec:conclusions} summarizes the main findings.

\section{Methods}
\label{sec:model}
This section presents the numerical models used in this work to simulate the electromagnetic behavior of the levitation systems. Section~\ref{sec:ja_resolved} describes the conventional approach to model these systems with the $j$-$a$ formulation. Section~\ref{sec:ja_homogenized} discusses applying homogenization for DCLs, and section~\ref{sec:ja_fixed_mesh} presents the approach to model the motion in the systems with a fixed mesh.

\subsection{Resolved model}
\label{sec:ja_resolved}
The FE models discussed in this work are based on the $j$-$a$ formulation but similar modeling can be carried out with other formulations. The $j$-$a$ formulation is an efficient choice for simulating the electromagnetic behavior of HTS devices, particularly in 2D~\cite{Stenvall_2010ab, dos-Santos_2023aa}. The approach expresses the material property in terms of the resistivity $\rho$ in the presence of the strong nonlinearity of the $e$-$j$ constitutive relation characteristic of superconductors~\cite{Dular_2020aa}. The computational domain $\Omega$ is divided into two distinct regions: a conducting region $\Oc$ (containing HTS) and a non-conducting region $\Occ$. Ampère's law expressed in terms of the magnetic vector potential $\a$ can be written as
\begin{equation}
\curl{\left( \nu \curl{\a} \right)} = \j,
\label{eq:ampere_A}
\end{equation}
where $\nu$ denotes the magnetic reluctivity and $\j$ the current density. The magnetic flux density $\b$ caused by permanent magnets (PMs) can be included by adding a field contribution $\curl{\left(\nu\b_{\mathrm{r}}\right)}$ to the right-hand side of \eqref{eq:ampere_A}, where $\b_{\mathrm{r}}$ is the remanent magnetic flux density. This term is added only to the regions containing permanent magnets $\Opm$. Ohm's law can be expanded to
\begin{equation}
\rho(|\j|) \,\j = -\partial_{t}\a - \grad{v},
\label{eq:J_ohms_law}
\end{equation}
where $v$ is the electric scalar potential. This resistivity is given by the power-law characteristic of high-temperature superconducting materials \cite{Rhyner_1993aa}
\begin{equation}
\rho(|\j|) = \frac{\ec}{\jc} \left( \frac{|\j|}{\jc} \right)^{\!n-1},
\label{eq:rho_power_law}
\end{equation}
with $\ec$ the critical electric field criterion, $\jc$ the critical current density, and $n$ the power-law exponent that governs the steepness of the superconducting-to-normal transition. For the parameters of the power law, different relations with or without field and temperature dependencies are used in practice to express $\jc$ and $n$. We briefly introduce the approximations used in the two separate simulation examples studied in this work. Exponential dependencies can be expressed as
\begin{equation}
    \jc = j_{\text{c,0}}\,\text{exp}\left(-\lvert\h\rvert/h_{\text{j}}\right)
    \label{eq:jc_verification}
\end{equation}
\begin{equation}
    n = n_{0}\,\text{exp}\left(-\lvert\h\rvert/h_{\text{n}}\right),
    \label{eq:n_verification}
\end{equation}
where $\h$ is the magnetic field strength, and $j_{\text{c,0}}$, $h_{\text{j}}$, $n_{0}$, and $h_{\text{n}}$ are the parameters of the material fit which are chosen as in \cite{Santos_2024aa}. A Kim-like approximation for $\jc$ is given as
\begin{equation}
    \jc = j_{\text{c,0}}\left(1+\frac{\sqrt{\left(kb_{\parallel}\right)^{2}+b_{\perp}^{2}}}{b_{0}}\right)^{-b_{\text{exp}}},
        \label{eq:jc_smb}
\end{equation}
where $j_{\text{c,0}}$, $b_{0}$,  $k$, and $b_{\text{exp}}$ are the parameters of the material fit chosen as in \cite{Santos_2023aa}, and $b_{\parallel}$ and $b_{\perp}$ are the magnetic flux density components parallel and perpendicular to the wide surface of the coated conductor, respectively.

Equations \eqref{eq:ampere_A} and \eqref{eq:J_ohms_law} need to be brought to a weak form for an approximate solution with FEM. Using the Galerkin method and introducing the function space $\a \in H(\text{curl}; \O)$, the weak form for \eqref{eq:ampere_A} including the field term from the PMs can be written as
\begin{equation}
\volInt{\nu\curl{\a}}{\curl{\a'}}{\O} - \surInt{\nu \curl{\a} \times \mathbf{n}}{\a'}{\partial\O} -  \volInt{\j}{\,\a'}{\Oc} = \volInt{\nu\b_{\mathrm{r}}}{\curl{\a}'}{\Opm}
\label{eq:weak_A}
\end{equation}
$\forall\a'$, where $\mathbf{n}$ is the outward unit normal vector on the boundary $\partial\O$. The volume integral over $\Omega$ and the surface integral over $\partial \Omega$ of the inner product of two fields are denoted by $\volInt{.\,}{\,.}{\Omega}$ and $\surInt{.\,}{\,.}{\partial \Omega}$, respectively. The surface integral term corresponds to the boundary conditions of the problem. In a similar manner, introducing $\j \in H(\text{div}; \Oc)$ allows writing the weak form for \eqref{eq:J_ohms_law} as
\begin{equation}
\volInt{\rho(|\j|) \, \j }{\j'}{\Oc} + \volInt{{\partial_{t} \a}}{\j'}{\Oc} + \volInt{\grad{v}}{\j'}{\Oc} = 0
\label{eq:weak_J}
\end{equation}
$\forall\j'$.
The nonlinear resistivity from \eqref{eq:rho_power_law} introduces material nonlinearity and is typically handled iteratively (e.g., via the Newton--Raphson method) at each time step of the transient simulation.

\begin{figure}
    \centering
    \begin{subfigure}[b]{0.45\linewidth}
        \centering
        \includegraphics[width=0.87\linewidth]{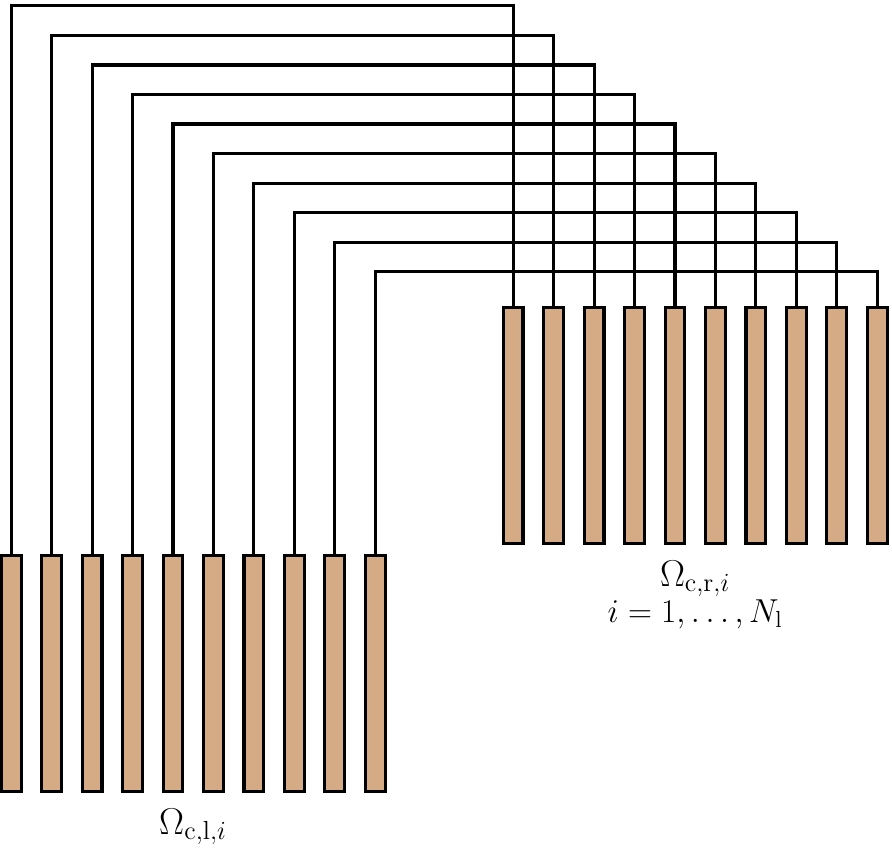}
        \caption{}
        \label{fig:dcl_homogenization_a}
    \end{subfigure}
    \begin{subfigure}[b]{0.45\linewidth}
        \centering
        \includegraphics[width=0.87\linewidth, trim={0em -2.6em 0em 0em}, clip]{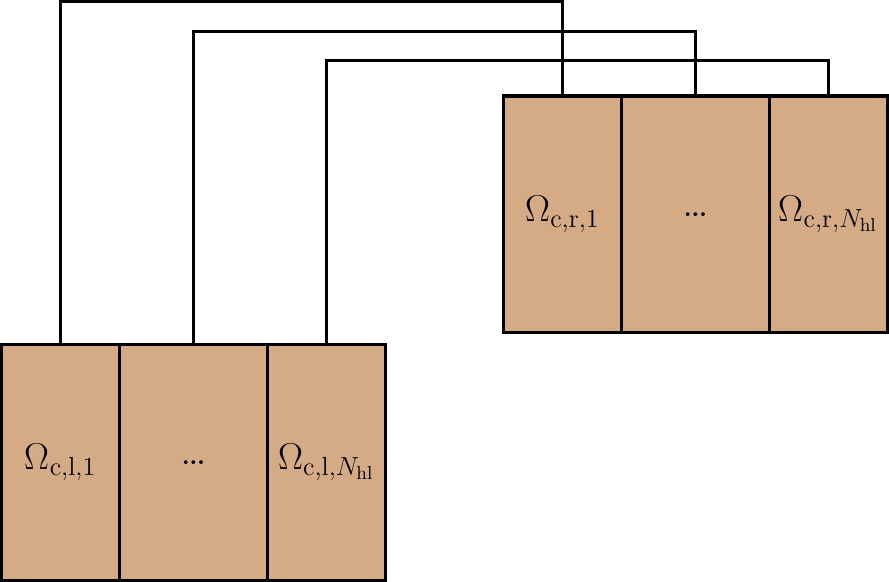}
        \caption{}
        \label{fig:dcl_homogenization_b}
    \end{subfigure}
    \caption{Illustration of a crossed loop cross-section (a) when all the CCs are resolved and (b) when homogenization is applied. The circuit coupling of the conducting domains is indicated for both cases.}
    \label{fig:dcl_homogenization}
\end{figure}

To enable simulations of practical devices driven by external power sources, the field problem has to be coupled to lumped-parameter electric circuits as is presented, e.g., for HTS tapes using the $j$-$a$ formulation in \cite{10493071}. Introducing $\grad v \in H(\text{curl}; \Oc)$, the global quantities relate to the local field quantities as~\cite{Dular_1999aa}
\begin{equation}
    -\volInt{\j}{\grad v'}{\Oci} = {I}_{i} {V}_{i}^{'}
    \label{eq:ja_wf2}
\end{equation}
$\forall v', V_{i}^{'}$, where $I_{i}$ and $V_{i}$ are the global currents and voltages of the $i$th separate conducting region in $\Oc$. A global basis function for each conducting region $\grad\vsi$ is defined as $\grad v = V_{i}\grad\vsi$. As field--circuit coupling is only needed in this work for the superconducting regions, in what follows $i=1,...,2\Nl$ for all of the $2\Nl$ separate superconducting regions. The loops formed from the slit CCs require an additional current constraint in 2D
\begin{equation}
    I_{k} = \int_{\OcSD{l}{k}} \jz\, d\OcSD{l}{k} = - \int_{\OcSD{r}{k}} \jz\, d\OcSD{r}{k}\, ,
    \label{eq:current_constraints}
\end{equation}
where $\jz$ is the current density perpendicular to the 2D plane, and $\OcSD{l}{k}$ and $\OcSD{r}{k}$ are the left and right conductor cross-sections of the $k$th loop, respectively. The loop geometry and the corresponding circuit constraints are illustrated in figure~\ref{fig:dcl_homogenization_a} for half of a DCL. With $2\Nl$ superconductor regions in the geometry which are all part of a loop, $\Nl$ loops are formed each needing one constraint. Together, the weak forms \eqref{eq:weak_A}--\eqref{eq:ja_wf2} and the current constraints \eqref{eq:current_constraints} form the $j$-$a$ formulated system examined in this work for the simulation of superconducting magnetic levitation systems.

The $j$-$a$ formulation is a mixed finite element formulation whose spatial discretization requires careful selection of the function spaces \cite{Dular_2021aa, Wang_2023aa}. Inadequate choices have been shown to lead to spurious oscillations in the obtained solution. The numerical implementation in this work follows closely that of \cite{Paakkunainen_2025aa}, where second order basis functions are used for $\a$ in $\Oc$ and first order basis functions are used for $\a$ elsewhere and for $\j$ and $\grad v$.

In a post-processing step, the force $\f$ induced on the HTS regions can be calculated as the Lorentz force
\begin{equation}
    \f = \int_{\Osc} \j \times \b\,\, \text{d}\Osc,
\end{equation}
where the region containing the superconductors $\Osc$ is introduced for convenience of notation, $\Osc \subset \Oc$.
To enable comparison of the force obtained from 2D simulations with the experimental results, the computed force is multiplied by the length of the HTS stacks, and end effects are taken into account with a scaling factor. To quantify the agreement between different results, the coefficient of determination $R^{2}$ for a force component $f_{\star}$ with respect to the reference $f_{\text{ref}}$ is calculated as
\begin{equation}
    R^{2} = 1 - \frac{\displaystyle\int_{0}^{T}\left(f_{\star}-f_{\text{ref}}\right)^{2}dt}{\displaystyle\int_{0}^{T}\left(f_{\text{ref}}-\bar{f}_{\text{ref}}\right)^{2}dt},
    \label{eq:r2}
\end{equation}
where $\bar{f}_{\text{ref}}$ is the mean value of the reference force over the time interval of the simulation $[0,\,T]$.

\subsection{Homogenized model}
\label{sec:ja_homogenized}
Homogenization approaches have been demonstrated to significantly reduce the computational cost of simulating systems with HTS while maintaining good accuracy \cite{Zermeno_2013aa, Berrospe-Juarez_2021aa}. Figure~\ref{fig:dcl_homogenization} shows the homogenization of a crossed loop where the $\Nl$ separate CC loops are replaced with a homogenized bulk material. The approach used closely resembles that presented in \cite{Zermeno_2013aa}, as the homogenized material is divided into $\Nhl$ homogenized loops (or $2\Nhl$ subdomains if following the naming convention of \cite{Zermeno_2013aa}) with $\Nhl < \Nl$, and each homogenized loop requiring a current constraint. The left and right sides of the homogenized loops are again connected through circuit coupling according to \eqref{eq:current_constraints}.

When the homogenized model in figure~\ref{fig:dcl_homogenization_b} is used, the material properties in $\Oc$ need to be replaced with effective material properties. In the model used here, only the HTS layer in each CC is taken into account. Under this approximation, the critical current density $\jc$ of the homogenized material can be defined according to the fill factor $\lambda$ of HTS in the CC as $\jc = \lambda \jchts$. With these modifications, the homogenized model is described by the same equations as the resolved model in section~\ref{sec:ja_resolved}, and, e.g., the forces inflicted on $\Oc$ can be computed in the same way as previously.

\subsection{Treatment of motion}
\label{sec:ja_fixed_mesh}
The FE model described so far does not explicitly include the relative motion of the model regions. Nevertheless, the model could capture the motion effects if a new mesh with updated positions is provided as the simulation progresses and the time derivative terms are treated correctly. In the following, we present an approach which requires no remeshing to capture the motion-induced effects. The rationale for the approach follows that presented in~\cite{Kettunen_2014aa}. In this work, the motion of the domains is prescribed, and no coupling to equations of motion is considered.

\begin{figure}
    \centering
    \includegraphics[width=0.5\linewidth]{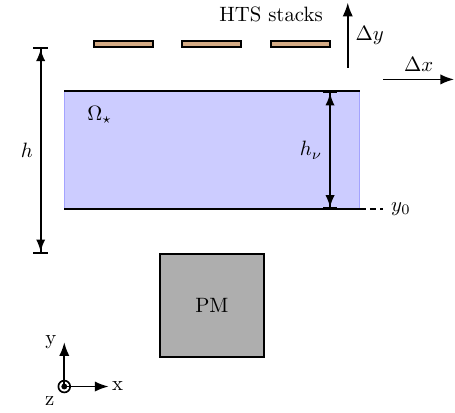}
    \caption{Sketch of a geometry containing HTS stacks and a PM. The domain $\Oairnu$ with the $\left(u,v,w\right)$ coordinate system is highlighted in blue. The positive directions of the displacements $\Delta x$ and $\Delta y$ are indicated with arrows.}
    \label{fig:verification_geometry}
\end{figure}

Let us examine a geometry with HTS stacks, a PM and the surrounding air, as illustrated in figure~\ref{fig:verification_geometry}. The nonconducting (air) domain $\Occ$ is divided into a conventional air domain $\Oair$ and into a domain $\Oairnu$ which we aim to utilize to model the effects of motion, $\Occ = \Oair \cup \Oairnu$. The domain $\Oairnu$ is highlighted in blue in figure~\ref{fig:verification_geometry}. The regions above and below $\Oairnu$ are assumed to be displaced by $\Delta x$ in the $x$-direction and $\Delta y$ in the $y$-direction with respect to each other. The translational motion can be expressed by introducing the coordinate system $\left(u,v,w\right)$ in $\Oairnu$. The transformation between the mesh coordinates $\left(x,y,z\right)$- and the $\left(u,v,w\right)$-coordinates is denoted as $\boldsymbol{\phi}:\Omega_{\text{xyz}} \to \Omega_{\text{uvw}}$ (the inverse transformation is $\boldsymbol{\phi}^{-1}: \Omega_{\text{uvw}} \to \Omega_{\text{xyz}}$) and is defined as
\begin{equation}
    \boldsymbol{\phi}(\mathbf{x}) = \mathbf{u} =
    \begin{bmatrix}
        x + \frac{y - y_{0}}{\hnu}\Delta x \\
        y + \frac{y - y_{0}}{\hnu}\Delta y \\
        z
    \end{bmatrix},
\end{equation}
where $\mathbf{x}$ is the coordinate vector $\left[x,y,z\right]^{\top}$, $\mathbf{u}$ the vector $\left[u,v,w\right]^{\top}$, $y_{0}$ the $y$-coordinate of the lower boundary of $\Oairnu$ and $\hnu$ the height of $\Oairnu$ in the mesh coordinates. The Jacobians of $\boldsymbol{\phi}$ and $\boldsymbol{\phi}^{-1}$ are denoted as $J_{\boldsymbol{\phi}}:=\text{d}\boldsymbol{\phi}/\text{d}\mathbf{x}$ and $J_{\boldsymbol{\phi}^{-1}}:=\text{d}\boldsymbol{\phi}^{-1}/\text{d}\mathbf{u}$, respectively. The magnetic flux density can be transformed between the $\left(u,v,w\right)$- and the $\left(x,y,z\right)$-coordinates as \cite{Hiptmair_2002aa}
\begin{equation}
    \b_{\text{uvw}} = \text{det}\left(J_{\boldsymbol{\phi}^{-1}}\right) J_{\boldsymbol{\phi}^{-1}}^{-1} \b,
    \label{eq:b_transform}
\end{equation}
where, to simplify the notation in what follows, only the $\left(u,v,w\right)$-coordinates are explicitly marked and the remaining field quantities are assumed to be in the mesh coordinates. The transformation \eqref{eq:b_transform} is directly introduced for brevity to avoid a derivation based on differential forms, and relies on the property that $\b$ is a 2-form. A similar problem containing motion is discussed in \cite{Kettunen_2014aa} using differential forms.

The presented model is modified only in $\Oairnu$ where now the curl-curl term in \eqref{eq:weak_A} holds in the $\left(u,v,w\right)$-coordinates. By using \eqref{eq:b_transform} and $\text{d}\O_{\text{uvw}} = 1 / \lvert\text{det}(J_{\boldsymbol{\phi}^{-1}})\rvert\,\text{d}\O$, the integral can be transformed back to the mesh coordinates as
\begin{align}
    &\int_{\Oairnu} \nu_{0}\,(\curl{\a})_{\text{uvw}}\cdot(\curl{\a'})_{\text{uvw}}\, \text{d}\O_{\text{uvw}} \nonumber \\
    =&\int_{\Oairnu} \nu_{0}\,\left(\text{det}\left(J_{\boldsymbol{\phi}^{-1}}\right) J_{\boldsymbol{\phi}^{-1}}^{-1} \curl{\a}\right)\cdot\left(\text{det}\left(J_{\boldsymbol{\phi}^{-1}}\right) J_{\boldsymbol{\phi}^{-1}}^{-1} \curl{\a'}\right)\,1 / \lvert\text{det}(J_{\boldsymbol{\phi}^{-1}})\rvert\,\text{d}\O \nonumber \\
    =&\int_{\Oairnu} \nu_{0}\,\text{det}\left(J_{\boldsymbol{\phi}^{-1}}\right) J_{\boldsymbol{\phi}^{-1}}^{-\top} J_{\boldsymbol{\phi}^{-1}}^{-1} \curl{\a}\cdot \curl{\a'}\,\text{d}\O,
    \label{eq:curlcurl_transform}
\end{align}
where $\nu_{0}$ is the reluctivity of free space.
The last line of \eqref{eq:curlcurl_transform} uses the condition $\hnu + \Delta y > 0$ in physically relevant cases to simplify the absolute value of the Jacobian determinant. A reluctivity tensor $\boldsymbol{\nu}$ can be defined in $\Oairnu$ as
\begin{equation}
    \boldsymbol{\nu} = \nu_{0}\,\text{det}\left(J_{\boldsymbol{\phi}^{-1}}\right) J_{\boldsymbol{\phi}^{-1}}^{-\top} J_{\boldsymbol{\phi}^{-1}}^{-1} =
    \nu_{0}\begin{bmatrix}
        \relcomp{xx} & \relcomp{xy} & 0 \\
        \relcomp{yx} & \relcomp{yy} & 0 \\
        0 & 0 & \relcomp{zz}
    \end{bmatrix},
    \label{eq:relutensor}
\end{equation}
with $\relcomp{xy} = \relcomp{yx}$, i.e., the tensor is symmetric. The coefficients read
\begin{align}
    \relcomp{xx} &= \frac{\hnu}{\hnu + \Delta y} \\[4pt]
    \relcomp{xy} &= \frac{\Delta x}{\hnu + \Delta y} \\[4pt]
    \relcomp{yy} &= \frac{\hnu + \Delta y}{\hnu} + \frac{\left(\Delta x\right)^{2}}{\hnu\left(\hnu + \Delta y\right)} \\[4pt]
    \relcomp{zz} &= \relcomp{xx}.
\end{align}
For motion only along the $y$-direction, the expression simplifies to $\boldsymbol{\nu} = \nu_{0} \, \text{diag}\left(\relcomp{xx},\, \relcomp{xx}^{-1},\, \relcomp{xx}\right)$. It is noted that the metric tensor $\g=J_{\boldsymbol{\phi}^{-1}}^{-\top} J_{\boldsymbol{\phi}^{-1}}^{-1}$ of the ($u,v,w$)-coordinate system emerges in~\eqref{eq:relutensor}. 

When the trajectory of the motion is known a priori, the movement can be implemented by a time-dependent $\boldsymbol{\nu}$. In the presented model, only the tensorial reluctivity in the curl-curl term $\volInt{\boldsymbol{\nu}\,\curl{\a}}{\curl{\a}'}{\Oairnu}$ needs to be introduced according to \eqref{eq:relutensor}. The rest of the formulation remains unchanged, and the same  mesh can be used for every time step in the simulation. It is noted that the transformation in general cases virtually deforms the elements in $\Oairnu$, meaning that despite the mesh staying the same, the effect on the field solution is the same as if the elements were stretched or compressed. This can affect the accuracy of the solution. For the translational motion examined in this work, the virtual deformations did not deteriorate the accuracy when using a reasonably fine structured rectangular mesh in $\Oairnu$. By contrast, \cite{Kettunen_2014aa} reports that large rotation angles reduce the solution accuracy.

Electromagnetic fields and Maxwell's equations can be expressed independently from the coordinate systems, or alternatively of the metric tensor \cite{Hiptmair_2002aa}. Consequently, a chosen coordinate system only affects the material properties which in this case is the relation between $\b$ and $\h$. The motion of the domains of the system which is expressed by a change of coordinates can then be incorporated into the constitutive relations. This gives a rough idea on the theoretical background why the motion in the system can be incorporated into the material properties. The approach is presented here for the $j$-$a$ formulation, but it can be applied to other formulations, e.g., to the $h$ formulation which is also commonly used for superconductors \cite{Dutoit_2023aa}. In this case, a similar derivation would yield a tensorial permeability $\boldsymbol{\mu}$ which could be used to capture the effect of the motion. Consequently, the approach is straightforward to implement in various existing simulation workflows.

\section{Numerical examples}
\label{sec:results}
This section presents numerical examples of the methods proposed in section~\ref{sec:model}. Section~\ref{sec:verification} shows a verification problem to verify and validate the fixed-mesh based approach, and section~\ref{sec:mb_simulation} demonstrates a more complex simulation of an SMB with the proposed methods. The results are compared with experimental data and with reference models from the literature. The models proposed in this paper are implemented using Gmsh~\cite{Geuzaine_2009ab} for mesh generation and GetDP~\cite{Dular_1998ac} for the FE solution. The code implementations are available in \cite{Paakkunainen_2026ae}.

\subsection{Verification and validation}
\label{sec:verification}

The verification example is taken from \cite{Santos_2024aa} and \cite{Sass_2015aa}. The problem is implemented using two approaches to take into account the motion. In the remeshed moving-geometry model, the geometry is updated and a new mesh is generated at each time step. In the fixed-mesh model, the meshed geometry remains unchanged, while the reluctivity tensor is updated according to \eqref{eq:relutensor}. Both of the models are implemented in GetDP, and the results are validated against measurements from the literature.

Figure~\ref{fig:verification_geometry} shows the geometry of the problem consisting of a PM and three separate, uncoupled, stacks of HTS CCs. The square shaped PM has the side length of \SI{50}{\milli\meter} and the remanent magnetization of \SI{780}{\kilo\ampere\per\meter} (or alternatively remanent flux density of approximately \SI{0.98}{\tesla}) in the positive $y$-direction. The HTS stacks are moved from the initial height $h=\SI{100}{\milli\meter}$ at a constant speed of \SI{0.5}{\milli\meter/\second} until $h=\SI{5}{\milli\meter}$, and then back to the initial height. The HTS stacks have been field cooled at the initial height. The PM material is assumed to be nonconducting and to have the relative permeability $\mu_{\text{r}} = 1$. Each HTS stack consists of 10 CCs, each \SI{12}{\milli\meter} wide and \SI{62.1}{\micro\meter} thick. The gaps between the stacks are \SI{1}{\milli\meter} in the $x$-direction. The separate stacks are homogenized with a single subdomain each, and the field dependent exponentional relations \eqref{eq:jc_verification} and \eqref{eq:n_verification}  are used for $\jc$ and $n$, respectively.

Standard meshing procedures are sufficient for both of the examined models with the mesh density chosen to achieve the desired accuracy. With the remeshed model, the mesh changes only along the path of the motion due to repeated generation of a triangular mesh. In other parts of the modeling domain, the mesh is kept unchanged to reduce noise in the field solutions caused by the varying mesh.

\begin{figure}
    \centering
    \includegraphics[width=0.8\textwidth]{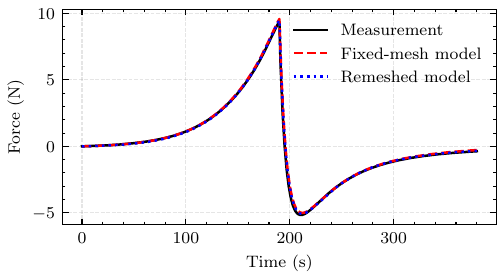}
    \caption{Comparison of the levitation force for the verification example. The simulated forces from the 2D models are scaled with a length of \SI{67}{\milli\meter} to allow for comparison with the measured force from~\cite{Sass_2015aa}. The maximum forces are for the measurements \SI{9.37}{\newton}, for the fixed-mesh model \SI{9.57}{\newton}, and for the remeshed model \SI{9.48}{\newton}.}
    \label{fig:verification_force}
\end{figure}

\begin{figure}
    \centering
    \includegraphics[width=0.95\textwidth]{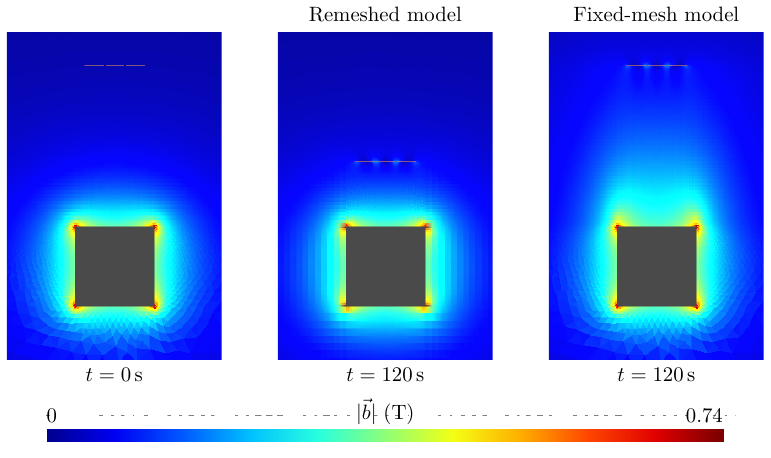}
    \caption{Magnitude of the magnetic flux density for the verification example with the remeshed and fixed-mesh models. Left: The initial configuration for both the models at $t=\SI{0}{\second}$. Middle: The remeshed model at $t=\SI{120}{\second}$. Right: The fixed-mesh model at $t=\SI{120}{\second}$. The field plots have been zoomed in to show the HTS stacks and the PM.}
    \label{fig:verification_b_field}
\end{figure}

Figure~\ref{fig:verification_force} shows the levitation force, i.e. the $y$-component of the force in the coordinate system of figure~\ref{fig:verification_geometry}, obtained from the different simulation approaches and measurements. The predicted and measured forces agree closely, thus validating the fixed-mesh model. To further quantify the agreement, the $R^2$ values are calculated with respect to the measurements for $f_{\text{y}}$ according to \eqref{eq:r2}. For the remeshed model, the coefficient is \SI{0.9983}{} and for the fixed-mesh model \SI{0.9975}{}. Figure~\ref{fig:verification_b_field} shows the magnitude of the magnetic flux density for both of the models at two different time instants. For $t=\SI{0}{\second}$, the initial conditions are the same, and thus also the field distributions apart from minor differences due to the different meshing. For $t=\SI{120}{\second}$, the HTS stacks have moved closer to the PM, as is seen for the remeshed model whereas the fixed-mesh model keeps the initial meshed geometry. However, with both models, the field distributions around the HTS stacks are very similar, demonstrating the capability of the fixed-mesh model to accurately capture the motion-induced effects. Attention needs to be paid in the post-processing phase to evaluate the magnitude of the magnetic flux density $\lvert \vec b \rvert$ correctly in $\Oairnu$ when using the mesh coordinates. By using the transformation~\eqref{eq:b_transform}, the norm can be expressed as
\begin{equation}
    \lvert \vec b \rvert = \sqrt{\vec b_{\text{uvw}}\cdot\vec b_{\text{uvw}}} = \text{det}\left(J_{\boldsymbol{\phi}^{-1}}\right) \sqrt{\vec b\cdot\left(\g\vec b\right)}.
\end{equation}

\begin{figure}
    \centering
    \includegraphics[width=0.8\textwidth]{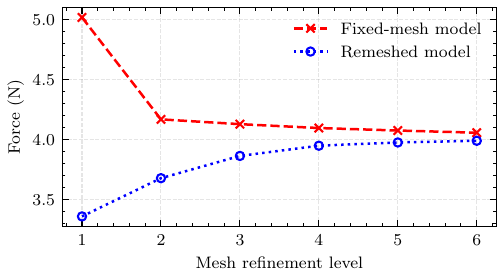}
    \caption{Mesh refinement study of the levitation force for the verification example. The force values are shown for the time $t=\SI{150}{\second}$. The number of degrees of freedom varies from \SI{3.5}{}k to \SI{28.4}{}k for the remeshed model and from \SI{3.3}{}k to \SI{24.8}{}k for the fixed-mesh model.}
    \label{fig:verification_convergence}
\end{figure}

Additionally, a mesh-refinement study is carried out for both of the models, and figure~\ref{fig:verification_convergence} shows the results for the levitation force at time $t=\SI{150}{\second}$. By refining the mesh, the results given by the models converge towards the same value. Apart from the coarsest mesh, the results with the fixed-mesh model change less with further mesh refinement. The results show that the fixed-mesh model is a valid approach to simulate the motion of superconductors in an external magnetic field.

\subsection{Magnetic bearing simulation}
\label{sec:mb_simulation}

Having verified and validated the fixed-mesh model, we examine a more complicated simulation of an SMB where additionally homogenization including multiple homogenized loops and circuit coupling are applied. The simulated SMB has been examined earlier in \cite{Martins_2019aa} and \cite{Santos_2023aa}, and the geometry is sketched in figure~\ref{fig:dcl_geometry}. The SMB is modeled in 2D by considering a cross-section of the geometry from the straight section of the DCLs. Measured results are compared with a reference model with a time-varying geometry and all of the conductors of the DCLs are resolved. The reference model is implemented with the $t$-$a$ formulation in COMSOL Multiphysics~\cite{Santos_2023aa}. The fixed-mesh model is implemented with GetDP.

The dimensions of the magnetic guideway are reported in \cite{Martins_2019aa}, and the geometry cross-section contains $\Nl = 100$ loops in total (25 loops in each crossed loop). The inner width of the DCLs is \SI{55}{\milli\meter}, and the two DCLs are \SI{10.4}{\milli\meter} apart from each other. The original unslit CCs are \SI{12}{\milli\meter} wide and their $\jc$ is described by \eqref{eq:jc_smb}. A constant power law exponent $n=20$ is used, and the fill factor of HTS is \SI{5.2e-3}{}. Both the PM and the iron in the magnetic guideway are assumed to be conducting with the values \SI{7.14e5}{\siemens/\meter} and \SI{8.41e6}{\siemens/\meter}, respectively. For the PMs, a remanent flux density of \SI{1.1}{\tesla} is given, and $\mu_{\text{r}}=1$. In contrast to \cite{Santos_2023aa}, for iron a constant $\mu_{\text{r}}=100$ is assumed for all the models to avoid having to export material functions between simulation software. Replacing the nonlinear $\b(\h)$ curve with a constant permeability is observed to have a negligible effect on the results.

\begin{figure}
    \centering
    \includegraphics[width=0.8\textwidth]{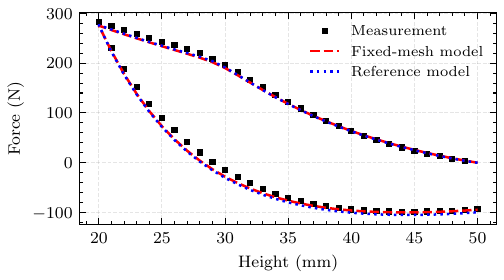}
    \caption{Vertical motion levitation force comparison with measurements from~\cite{Martins_2019aa} and different model implementations. The maximum forces are for the measurements \SI{282}{\newton}, for the reference model \SI{276}{\newton}, and for the fixed-mesh model \SI{275}{\newton}.}
    \label{fig:mb_fy}
\end{figure}

We examine the forces acting on the SMB when the DCLs are at different positions with respect to the magnetic guideway. The SMB cross-section uses the same coordinate system as figure~\ref{fig:verification_geometry}. In all of the examined cases, the DCLs are initially at the height of $h=\SI{50}{\milli\meter}$ above the guideway. The DCLs have been field cooled at this height. First, we examine the levitation force $\fy$ when the motion is in the $y$-direction. The DCLs are moved from $h=\SI{50}{\milli\meter}$ to $h=\SI{20}{\milli\meter}$, and then back to $h=\SI{50}{\milli\meter}$ at a constant speed of \SI{1}{\milli\meter/\second}. Figure~\ref{fig:mb_fy} compares the experimental data with results from the reference and fixed-mesh models. The results show good agreement of levitation forces. Again, the $R^2$ values are calculated with respect to the measurements, and values of \SI{0.9947}{} and \SI{0.9960}{} are obtained for the reference and the fixed-mesh model, respectively. To allow for comparison of experimental results and the 2D simulation, a length of \SI{100}{\milli\meter} is set for the SMB geometry, and an additional scaling factor of 1.4 is introduced to account for end effects \cite{Santos_2023aa}.

\begin{figure}
    \centering
    \includegraphics[width=0.9\textwidth]{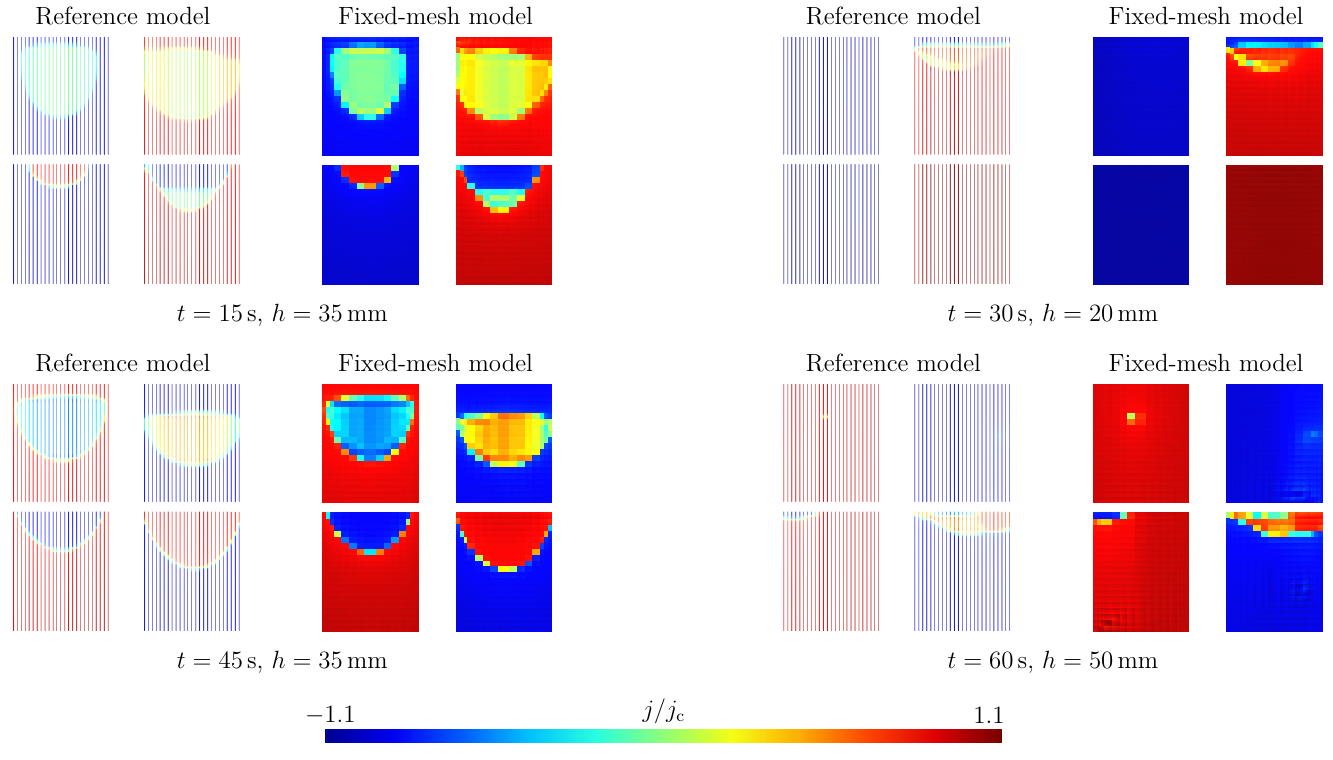}
    \caption{Current density distributions normalized with respect to $\jc$ for the reference and fixed-mesh models during the vertical motion. Due to symmetry, only one of the DCLs is shown.}
    \label{fig:mb_j_field}
\end{figure}

Figure~\ref{fig:mb_j_field} shows the comparison of the current density distributions normalized with respect to $\jc$ during the vertical motion. The distributions are shown for one of the DCLs in the SMB as the distributions are symmetric for the vertical motion, as shown in \cite{Santos_2023aa}. A very good agreement is observed between the reference and the fixed-mesh model for different positions along the trajectory.  

\begin{figure}
    \centering
    \begin{subfigure}[b]{0.5\linewidth}
        \hspace{-3em}
        \centering
        \includegraphics[width=0.95\linewidth]{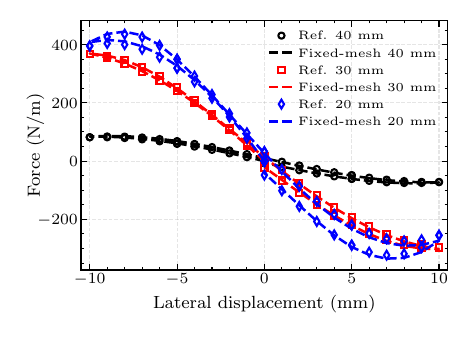}
        \caption{Computed $\fx$ values.}
        \label{fig:mb_fx_a}
    \end{subfigure}%
    \begin{subfigure}[b]{0.5\linewidth}
        \centering
        \includegraphics[width=0.95\linewidth]{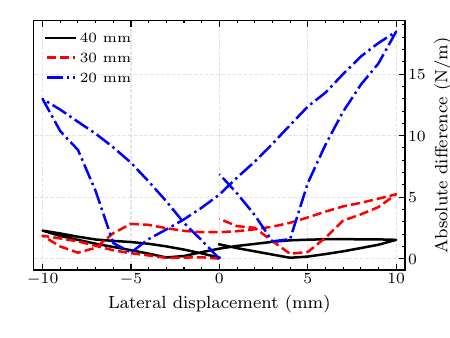}
        \caption{Absolute differences in $\fx$.}
        \label{fig:mb_fx_b}
    \end{subfigure}
    \caption{Comparison of lateral force per unit length between the different models over the cyclic horizontal motion: (a) forces obtained with the reference and the fixed-mesh model and (b) the corresponding absolute differences.}
    \label{fig:mb_fx}
\end{figure}

To demonstrate a more general motion, we examine the lateral force per unit length $\fx$, when the DCLs are first moved closer to the guideway with a vertical motion.  Then, a horizontal motion is carried out with again the constant velocity of \SI{1}{\milli\meter/\second} first until $\Delta x = \SI{-10}{\milli\meter}$, then until $\Delta x = \SI{10}{\milli\meter}$, and returning to the initial position $\Delta x = \SI{0}{}$. Figure~\ref{fig:mb_fx_a} shows $\fx$ during the motion for three different heights for the reference and the fixed-mesh model. The corresponding absolute differences between the models are shown in figure~\ref{fig:mb_fx_b}. The models are in good agreement with each other.

Lastly, the computational effort of the models is compared in the simulation case with lateral motion at the height of \SI{20}{\milli\meter}. Table~\ref{tab:levmb_numbers} contains the number of degrees of freedom (DoFs), $n_{\text{DoFs}}$, the number of linear systems solved, $n_{\text{linsys}}$, and the total simulation time, $t_{\text{tot}}$, for the models. Additionally, the effect of the total number of homogenized loops in the geometry, $\Nhl$, on the accuracy of the fixed-mesh model is examined. The accuracy is quantified with the $R^2$ values and the maximum absolute difference $\errmax$ which are calculated with respect to the reference model. Apart from $\Nhl$, the meshing parameters are kept the same. With higher $\Nhl$, the accuracy improves at the cost of increasing $n_{\text{DoFs}}$ and $t_{\text{tot}}$. Consistently, the fixed-mesh model provides a substantial reduction in runtime and in $n_{\text{DoFs}}$ while maintaining good accuracy as indicated by the $R^2$ and $\errmax$ values. When expressing the maximum absolute differences as corresponding relative differences, the values are \SI{10.7}{\percent}, \SI{7.2}{\percent}, and \SI{5.5}{\percent} for $\Nhl=\{12,20,28\}$, respectively. It is noted that a rigorous comparison of the simulation times across different FE software is difficult to establish, but still the results clearly indicate a speed-up of the simulations with the fixed-mesh model. For example, due to the used computing environment, the reference model runs in parallel across multiple processor cores, while the fixed-mesh model runs on a single core even though parallelization of the computation would be possible with GetDP.

\begin{table}
    \caption{Comparison of simulation models for the SMB example with lateral motion at \SI{20}{\milli\meter} height. The simulations were performed on a desktop computer with an Intel Xeon E5-2609 v3 CPU.}
    \centering
    \renewcommand{\arraystretch}{1.1} 
    \begin{tabular}{*{7}{c}}
    \br
    Model & $\Nhl$ & $R^2$ & $\errmax$ (\SI{}{\newton/\meter}) & $n_{\text{DoFs}}$ & $n_{\text{linsys}}$ & $t_{\text{tot}}$ \\
    \hline Reference & - & - & - & \SI{86.3}{}k$\,-\,$\SI{88.9}{}k\setcounter{footnote}{6}\footnotemark & \SI{6513}{} & \SI{8.35}{\hour} \\
    \hline \multirow{3}{*}{Fixed-mesh}
        & 12 & \SI{0.9976}{} & \SI{27.1}{} & \SI{18.7}{}k & \SI{1296}{} & \SI{0.26}{\hour} \\
        & 20 & \SI{0.9988}{} & \SI{18.5}{} & \SI{20.9}{}k & \SI{1427}{} & \SI{0.34}{\hour} \\
        & 28 & \SI{0.9993}{} & \SI{14.1}{} & \SI{22.3}{}k & \SI{1504}{} & \SI{0.40}{\hour} \\
    \br 
    \end{tabular}
    \label{tab:levmb_numbers}
\end{table}
\setcounter{footnote}{7}%
\footnotetext{$n_{\text{DoFs}}$ varies slightly for the reference model due to the remeshing during the motion.}

\section{Conclusions}
\label{sec:conclusions}
A fixed-mesh based modeling approach is presented for the simulation of translational motion in SMBs. The motion induced effects are modeled through a coordinate transformation applied only to an air region between the moving parts. The relative displacement is represented by a time-dependent anisotropic reluctivity tensor while the finite element mesh remains unchanged. The approach is implemented in 2D with FEM using the $j$-$a$ formulation, and is combined with homogenization and circuit coupling for DCLs. The approach requires only the implementation of the material tensor, and can be applied to existing simulation workflows and to other field formulations.

The approach is verified and validated against measurements and existing modeling approaches. In the verification example, the fixed-mesh model is in close agreement with both measurements and a remeshed model, achieving $R^2$ values above \SI{0.99}{} for the levitation force. The different models converge towards the same force under mesh refinement. In the SMB simulation, the fixed-mesh model reproduces the measured and previously simulated vertical force with high accuracy. Over the combined vertical and horizontal displacements, the homogenized fixed-mesh model is able to reproduce the results of the reference model which resolves all the CCs of the DCLs and accounts for the motion by updating the mesh during the simulation. Increasing the number of homogenized loops improves the accuracy of the homogenized model. Additionally, the fixed-mesh model is computationally more efficient than the reference model, by indicating a considerable reduction in the number of DoFs and a reduction of the simulation time from over \SI{8}{\hour} to under \SI{0.5}{\hour}.

The results demonstrate that the fixed-mesh approach can accurately and efficiently represent horizontal and vertical translations in SMBs. The carried out work provides clear directions for future research, including coupling the model to the equations of motion, considering more general motion, and modeling 3D systems.

\section*{Acknowledgments}
The work of Elias Paakkunainen is supported by the Graduate School CE within Computational Engineering at the Technical University of Darmstadt. Bárbara Maria Oliveira Santos and Gabriel dos Santos acknowledge the support of CAPES and CNPq - this work was financed in part by the Coordenação de Pessoal de Nível Superior (CAPES) - finance code 001, and by the Conselho Nacional de Desenvolvimento Científico e Tecnológico (CNPq).

\printbibliography

\end{document}